\documentclass[times, twoside, watermark]{open_source}
\usepackage{blindtext}

\leadauthor{Westner}

\begin{document}

\title{Cycling on the Freeway: The Perilous State of Open Source Neuroscience Software}
\shorttitle{}

\author[1,2,\text{*}\Letter]{Britta U. Westner}
\author[3,\text{*}]{Daniel R. McCloy}
\author[3,\text{*}]{Eric Larson}
\author[4]{Alexandre Gramfort}
\author[5]{Daniel S. Katz}
\author[6]{Arfon M. Smith}
\author[ ]{invited co-signees}

\affil[1]{Department of Cognitive Neuroscience, Donders Institute for Brain, Cognition and Behaviour, Radboud University Medical Center, Nijmegen, The Netherlands}
\affil[2]{Donders Centre for Cognition, Donders Institute for Brain, Cognition and Behaviour, Radboud University, Nijmegen, The Netherlands}
\affil[3]{Institute for Learning and Brain Sciences, University of Washington, Seattle, WA, United States}
\affil[4]{Université Paris-Saclay, Inria, CEA, Palaiseau, France}
\affil[5]{NCSA \& CS \& ECE \& iSchool, University of Illinois Urbana-Champaign, Illinois, United States}
\affil[6]{GitHub Inc.}
\affil[*]{\textit{Lead authors}}

\maketitle

\begin{abstract}
Most scientists need software to perform their research \citep{barker2020,carver2022,hettrick2014s,hettrick2014,switters2019}, and neuroscientists are no exception. Whether we work with reaction times, electrophysiological signals, or magnetic resonance imaging data, we rely on software to acquire, analyze, and statistically evaluate the raw data we obtain — or to generate such data if we work with simulations. In recent years there has been a shift toward relying on free, open-source scientific software (FOSSS) for neuroscience data analysis \citep{poldrack2019}, in line with the broader open science movement in academia \citep{mckiernan2016} and wider industry trends \citep{eghbal2016}. Importantly, FOSSS is typically developed by working scientists (not professional software developers) which sets up a precarious situation given the nature of the typical academic workplace (wherein academics, especially in their early careers, are on short and fixed term contracts). In this paper, we will argue that the existing ecosystem of neuroscientific open source software is brittle, and discuss why and how the neuroscience community needs to come together to ensure a healthy growth of our software landscape to the benefit of all.

\end {abstract}

\begin{keywords}
open science | open source | neuroscience | electrophysiology
\end{keywords}

\begin{corrauthor}
britta.westner[at]donders.ru.nl
\end{corrauthor}

\paragraph*{The neuroscience open source system is moving towards a crisis.}
The development and especially maintenance of open source scientific software is labor-intensive, especially when adopting best practices of open science: readability, resilience, and re-use \citep{connolly2023}. Critically, the survival of any open source software is guaranteed not only by the total number of contributors, but by the number of developers who have a sufficiently complete overview of the source code and the tools and processes in place for maintenance of the project. The \emph{Truck Factor} (or \emph{Bus Factor}) quantifies the number of such maintainers for a given project, and can be seen as an indicator of the risk of incomplete knowledge- and capability-sharing among the project’s team members \citep{avelino2016}. Table~\ref{tab:trucks} shows the Truck Factors of several widely used analysis software projects in electrophysiological neuroscience. It is evident that projects typically have a Truck Factor of one to three, meaning that only one to three people per project have sufficient knowledge to keep the project alive (note that the Truck Factor does not take into account whether those contributors with the relevant knowledge are still active or have funding/support to do the maintenance work). This reveals the fragility of the academic open source system: \emph{all users} of a software package rely on the work of \emph{one to three people} to keep everyone’s data analyses from failing. It also highlights the enormous pressure the maintainers of open source software are under.
Given the importance of scientific software to the practice of modern science, one would hope developing FOSSS to be a well-supported and incentivized role in academia. This hope is not borne out: the academic incentive structure for software development and maintenance is woefully inadequate \citep{carver2022,davenport2020,merow2023,smith2018} and can even be perceived as hostile \citep{millman2018,perez2011}. As \citet{davenport2020} state: ``In spite of the vital role research software plays, it largely remains undervalued, with time spent in training or development seen as detracting from the `real research'.'' The current incentive model often summarized as ``publish-or-perish'' has a significant negative impact on the software ecosystem’s health in several ways, which we will describe below.

\paragraph*{Rewarding publications in promotion disfavors software development.}
Many scientists who develop software publish a paper to introduce their project to the community, and direct users to cite that paper if they use the software for their analysis \citep{katz2018,smith2016}. However, software papers are usually published to mark the first public release or sometimes a later major release of a software package. Given that publication counts are such an important (and often the primary) metric for academic tenure and promotion, once the paper announcing the software is published, there is little incentive from the employer to continue developing the software. This may be changing as academic institutions rely more on composite citation metrics rather than simple publication counts — by incentivizing that the paper be widely cited — but the reward of new projects and attendant new publications can easily outweigh the reward of garnering more software-paper citations, especially given that software-papers are viewed as less valuable than research papers \citep{davenport2020}.
At the same time, people who contribute to development or maintenance of FOSSS at a later stage do not get rewarded by citations of the initial paper \citep{davenport2020}, thus increasing the chance that the software will become unmaintained. We worry that the lack of benefit for project ``latecomers'' also often incentivizes starting new projects (instead of contributing to existing ones) — a new project means a new publication! — which in turn risks further increasing the number of unmaintained software packages in the field.
To spell it out: open source work done by academics (in time they might otherwise use to do research) sustains \emph{other} academics writing \emph{their} papers \citep{merow2023} — which puts the software-developing academics behind in the publish-or-perish culture of academia.

\paragraph*{Lack of stable long-term funding for scientific software stifles its growth.}
Funding is hard to obtain for the development of new software — and even harder for the maintenance of existing software \citep{davenport2020}. That means that many contributors and maintainers either do not have secure long-term academic positions or are not paid primarily for their software work. This is slowly changing: programs like POSE from the US National Science Foundation, open science supplements to grants from the US National Institutes of Health, the UK Research and Innovation office, the national plan for open science (``plan national pour la science ouverte'') of the French government, and software-focused grants from organizations like CZI, the Sloan Foundation, and the Simons Foundation are all helping to support the development of FOSSS. But the (mostly) short-term grant-based nature of these new support mechanisms means that once the grant is over the primary incentive to continue maintaining the software is largely gone, a void quickly filled by the ever-present publish-or-perish incentive lurking in the background.

\paragraph*{Existing barriers skew the developer demographic to the detriment of science.} Research shows that diversity has positive effects on project outcomes generally \citep{earley2000,hoogendoorn2013,jackson2004,roberson2019} and for open source projects in particular \citep{daniel2013,vasilescu2015}. However, structural factors such as misogyny and racism hinder diversity of the developer pool: only 1.1-5.4\% of open source developers are perceptible as or identify as women \citep{eghbal2016,geiger2022,ghosh2002,nafus2012} and less than 17\% are perceptible as Non-White \citep{nadri2021}. The cost of contributing is higher for minorities \citep{whitaker2020}: not only do they often feel a pressure to have to be perfect \citep{singh2021}, underrepresented groups have to face stereotyping, discrimination, and harassment in the open source software world \citep{frluckaj2022,nadri2021,nafus2012,singh2021,vasilescu2015a}.

We are unaware of any systematic studies examining diversity in neuroscience software, but our collective experience in the field suggests that it is not substantially different from the broader open-source software community.  If true, this is unsurprising: academics need a certain level of privilege to be able to contribute to open source in the first place. Typical barriers to participation in FOSSS projects are the lack of permission and support from a supervisor, the lack of ``free'' time outside normal work hours due to family or financial demands on that time, and lack of confidence due to inadequate training, role models, and guidance. These barriers all affect marginalized groups more strongly, and are compounded by the lack of representation and role models from underrepresented demographics and by the myth of meritocracy \citep{nafus2012}. Moreover, the ``informalization'' of open source \citep{nafus2012} compounds the problem by eschewing traditional application and advancement processes and legal workplace protections in the spaces and interactions where FOSSS work is carried out, and consequently such spaces are often dominated by ``old boys'' networks where again underrepresented identities face an uphill battle. Moreover, the publish-or-perish culture and lack of stable funding for software development in academia also contribute to the substantial lack of qualified labor in FOSSS communities (cf. Table~\ref{tab:trucks}). This further impacts the diversity of FOSSS communities with respect to career stage (an imperfect proxy for age) — many academic open-source contributors are pursuing or have just finished their PhD and the devaluing of software work makes it harder for them to achieve tenured positions, while more senior academics who contributed early in their career are often not able to prioritize open source work anymore.

In summary, workplace incentives in academia disfavor participation in open source communities; the lack of a level playing field further discourages participation for certain classes of academics; and changing the incentives through extramural funding has so far been a temporary, partial fix on too small a scale to be transformative. But given the shift toward relying on FOSSS in neuroscience, it seems clear that the incentive structure must change. In the following, we will discuss ways in which such change can be realized.

\begin{table*}[t]
    \centering
    \begin{tabular}{lrl}
        \textbf{Software} & \textbf{Truck Factor} & \textbf{Hash} \\
         Brainstorm & 1 & \texttt{053b2ea6bb46ca633b931d9467f292bd7eae1361} \\
         EEGLAB & 1 & \texttt{cc87c2e08716d397da76821f0fd651bbb9c4357c}\\
         FieldTrip & 1 & \texttt{f41a2f7aa84d2fbebbf17369e155855b4358c3b8} \\
         MNE-Python & 3 & \texttt{9627f43bb2c03233827046bc06acdab0968d6610}\\
         SPM & 3 & \texttt{fba4f8e139c3f975b61e2581828d151b26dbe68a} \\
         Astropy & 6 & \texttt{53188c39a23c33b72df5850ec59e31886f84e29d} \\
    \end{tabular}
    \caption{The reported Truck Factors were estimated using the heuristic-based approach reported in \citet{avelino2016}. This algorithm estimated the Truck Factor using a degree-of-authorship metric \citep{fritz2010}: ``authors'' of files are those developers who are able to maintain a file moving forward. Developers who have (joint) authorship of at least 50\% of files count towards the Truck factor. For each package, the Truck Factor analysis was performed 29 August 2023; the commit hash for the state of the codebase when we performed the analysis is shown. We compare neuroscience packages to Astropy \citep{theastropycollaboration2013}, a well-supported and widely used software package in astronomy and astrophysics.
}    \label{tab:trucks}
\end{table*}

\paragraph*{Professionalize academic software development.} One obvious cornerstone for a more sustainable and professional FOSSS ecosystem is funding, especially funding that supports maintenance of existing, widely-used projects \citep{merow2023}. That funding should support scientists who engage in software development as part of their normal academic appointment duties. It has further been suggested that all levels of software development in academia (from widely-used packages down to single-user analysis scripts) can be carried out or supported by experienced research software engineers \citep{connolly2023,merow2023}. As others have noted, there are at least two obstacles to academic software work being carried out by professional software engineers: first, academia struggles to attract technically skilled professionals given the (much) higher compensation available in industry \citep{connolly2023,gewin2022,seidl2016}, and second, there is a perception among professional engineers that career advancement prospects in academia are limited \citep{carver2022,connolly2023,merow2023}.
However, even if institutional support for research software engineering were to increase dramatically, we see a further obstacle to the sustainability of FOSSS: academic software development needs the participation of scientists that are active in the field and have the domain knowledge necessary to understand relevant use cases, best practices, and common analysis pitfalls. This latter point also underscores the importance of having more senior academics involved in software development, which (as discussed above) is disincentivized and consequently rare. Thus we believe that a robust solution must involve institutions increasing their support of \emph{working scientists} who develop software (perhaps alongside professional research software engineers), and recognizing and rewarding software contributions as an integral part of the practice of modern science.

\paragraph*{Adopting new software citation practices can increase the valuation of FOSSS.} Increasing the valuation of open source work for academic career advancement is crucial. One suggested shortcut to this would be the exploitation of the existing publish-or-perish culture by introducing ``update publications'' after software development milestones \citep{merow2023}. However, in our view this risks perpetually re-creating the same set of problems, just on a shorter time-scale. Encouraging researchers to cite the software they use — and to cite the \emph{software itself} (i.e., the specific version used), not the canonical paper that introduced the software — seems like a better adaptation within the existing culture: it does not require the developers to write a new paper, and all contributors to a specific version of the software get credit. Indeed, ``cite the software, not the paper'' is considered best practice already \citep{katz2020,smith2016}, and we hope that this can become standard practice in neuroscience; publishers, journals, editors, and reviewers must play a role here in insisting that authors attribute their software usage in line with those best practices.

\paragraph*{Start viewing research software work as an important contribution to science.} We urge funding bodies and promotion committees to value substantial open source software work in their evaluation guidelines as an important contribution to science of equal value as one or (ideally) multiple  papers. The development of widely used analysis software should be acknowledged and rewarded as a contribution to science rather than viewed as merely the development of one’s coding skills. Individual investigators can play an important role here too: if your lab relies on FOSSS for data analysis, consider allocating grant funds to support a scientific software developer, making FOSSS contribution part of the job description for your next postdoc or research scientist hire, or planning that your graduate students will need either a longer duration of support or fewer research output expectations (or both) in order to develop the necessary competencies \emph{to both use and contribute to} the tools they rely on. Importantly, it should not be the developer’s responsibility to prove the value of their contributions; the burden should lie with promotion and tenure committees to become familiar with how to evaluate the scope and import of software contributions (fortunately, the Research Software Alliance and Research Data Alliance are working on policy recommendations on this topic,\footnote{https://www.rd-alliance.org/groups/policies-research-organisations-research-software-pro4rs} and OSPO++ is working to create, support and improve open-source program offices across academia\footnote{https://ospoplusplus.org/}). Failing that, promotion and tenure committees should at minimum allow, request, and encourage scientists who develop software to contextualize the scope and impact of their software work \citep{hafer2009}.
We however want to caution against the development of new and too simplistic metrics for the measure of open source work in academia, as measuring software impact is difficult and takes time \citep{afiaz2023}. Furthermore, Goodhart’s law postulates that every measure which becomes a target becomes a bad measure \citep{afiaz2023,goodhart1984}, as has happened to the \emph{h-index} \citep[e.g.,][]{bartneck2011,purvis2006,seeber2019,zhivotovsky2008}.

\paragraph*{Better training can facilitate a healthy FOSSS ecosystem in the long-term. } Lastly, we advocate for programs and departments to incorporate better training of students and PhD candidates in software development and maintenance \citep{carver2022,millman2018,guest2023}: this would not only increase the software engineering skills of academics generally  \citep{connolly2023}, but would also send a strong signal about the importance of well-written, reusable code to junior researchers. While training programs such as e.g., Software Carpentry\footnote{https://software-carpentry.org/}, CodeRefinery\footnote{https://coderefinery.org/}, INTERSECT\footnote{https://intersect-training.github.io}, Neurohackademy\footnote{https://neurohackademy.org} or Neuromatch Academy\footnote{https://academy.neuromatch.io} give excellent training courses and summer schools, we also call on teaching coordinators of universities to reflect an increasing need for programming literacy in their neuroscience programs. Together with the recognition for writing open source software, this would be an important step towards a healthy software ecosystem in neuroscience.

\paragraph*{A higher valuation of FOSSS will increase the inclusivity of open source spaces.} We predict that all action points discussed above will have a positive impact on the inclusivity of open source in academia. Changing the incentive structure such that open source work does not have to be a privilege anymore but gets seen as what it is: a valid and critical contribution to science, will facilitate the participation of underrepresented groups. Beyond this, we strongly advocate that open source projects in neuroscience make sure to be welcoming to everyone and to prevent any harassment, e.g. by stating and adhering to Community Guidelines.

\paragraph*{Conclusion} Summarizing the key points of this paper, we hope to raise awareness within the neuroscientific community about its dependence on a rather brittle structure. Your open source software ecosystem needs your help! Immediate action can be taken by citing current software versions instead of the seminal software-describing paper, by making space for your trainees to engage in FOSSS communities, and rewarding them (or at least not penalizing them) when they do. Beyond this, the incentive structure in academia and the policies that support it (including those created by research performing organizations, funders, publishers, etc.) urgently need to be re-thought \citep{hostler2023,jensen2023,merow2023,millman2018,munafo2017,neylon2012} — not only for the sake of the academic open source ecosystem, but for the good of the neuroscience community as a whole.

\begin{acknowledgements}
This project has been made possible in part by grant number 2021-237679 from the Chan Zuckerberg Initiative DAF, an advised fund of Silicon Valley Community Foundation.
\end{acknowledgements}

\section*{References}
\bibliography{open_source-zotero}

\end{document}